\documentclass[aip,apl,reprint,superscriptaddress,showpacs,amsmath,citeautoscript]{revtex4-1}
\usepackage{graphicx}
\usepackage{ucs}\usepackage[utf8x]{inputenc}
\usepackage{CJK}
\usepackage[slantedGreek]{mathptmx}\usepackage[scaled=.9]{helvet}\usepackage{courier}
\usepackage{mathcomp}
\usepackage[colorlinks,citecolor=blue,filecolor=blue,linkcolor=blue,urlcolor=blue]{hyperref}

\begin{document}

\title{The catalytic potential of high-$\kappa$ dielectrics for graphene formation}

\author{Andrew Scott}
\affiliation{Leibniz-Institut f\"{u}r Festk\"{o}rper- und Werkstoffforschung Dresden e.\,V., PF 27\,01\,16, 01171 Dresden, Germany}
\affiliation{Technische Universit\"{a}t Dresden, 01062 Dresden, Germany}
\author{Arezoo Dianat}
\affiliation{Technische Universit\"{a}t Dresden, 01062 Dresden, Germany}
\author{Felix B\"{o}rrnert}
\email{f.boerrnert@ifw-dresden.de}
\affiliation{Leibniz-Institut f\"{u}r Festk\"{o}rper- und Werkstoffforschung Dresden e.\,V., PF 27\,01\,16, 01171 Dresden, Germany}
\author{Alicja Bachmatiuk}
\affiliation{Leibniz-Institut f\"{u}r Festk\"{o}rper- und Werkstoffforschung Dresden e.\,V., PF 27\,01\,16, 01171 Dresden, Germany}
\author{Shasha Zhang (\begin{CJK*}{UTF8}{}\CJKfamily{gbsn}张莎莎\end{CJK*})}
\affiliation{Leibniz-Institut f\"{u}r Festk\"{o}rper- und Werkstoffforschung Dresden e.\,V., PF 27\,01\,16, 01171 Dresden, Germany}
\author{Jamie H. Warner}
\affiliation{University of Oxford, Parks Road, Oxford OX1 3PH, United Kingdom}
\author{Ewa Borowiak-Pale\'{n}}
\affiliation{Zachodniopomorski Uniwersytet Technologiczny, \mbox{Pulaskiego 10, 70322 Szczecin, Poland}}
\author{Martin Knupfer}
\affiliation{Leibniz-Institut f\"{u}r Festk\"{o}rper- und Werkstoffforschung Dresden e.\,V., PF 27\,01\,16, 01171 Dresden, Germany}
\author{Bernd B\"{u}chner}
\affiliation{Leibniz-Institut f\"{u}r Festk\"{o}rper- und Werkstoffforschung Dresden e.\,V., PF 27\,01\,16, 01171 Dresden, Germany}
\author{Gianaurelio Cuniberti}
\affiliation{Technische Universit\"{a}t Dresden, 01062 Dresden, Germany}
\affiliation{Division of IT Convergence Engineering, POSTECH, Pohang 790-784, Republic of Korea}
\author{Mark H. R\"{u}mmeli}
\email{m.ruemmeli@ifw-dresden.de}
\affiliation{Leibniz-Institut f\"{u}r Festk\"{o}rper- und Werkstoffforschung Dresden e.\,V., PF 27\,01\,16, 01171 Dresden, Germany}
\affiliation{Technische Universit\"{a}t Dresden, 01062 Dresden, Germany}

\date{\today}


\begin{abstract}
The growth of single and multilayer graphene nano-flakes on MgO and ZrO$_2$ at low temperatures is shown through transmission electron microscopy. The graphene nano-flakes are ubiquitously anchored at step edges on MgO (100) surfaces. Density functional theory investigations on MgO (100) indicate C$_2$H$_2$ decomposition and carbon adsorption at step-edges. Hence, both the experimental and theoretical data highlight the importance of step sites for graphene growth on MgO.
\end{abstract}

\maketitle

Graphene promises a great leap forward in electronic device speed due to its high charge carrier mobility \cite{geim09}. Restricting its dimensions sufficiently ($< 10$\:nm for room temperature applications) to introduce finite size effects can open an energy gap \cite{son06,shemella07}. Confining the width and keeping the length infinitely long forms a graphene nano-ribbon (GNR) whilst restricting both the width and length forms a graphene nano-flake. Interesting electronic properties can also arise from bi-layer nano-ribbons/flakes and even higher order stacked flakes. Their energy gaps are not just dependent on size, but also on the edge states and electric fields \cite{sahu08,lima09,sahu10a,sahu10b,xia10}. Hence, the richness of electronic properties afforded by graphene causes great excitement for its use in nano-electronic devices. However, various technical hurdles exist. For example, the implementation of graphene in nano-electronic technology is hindered by difficulties in obtaining defect free and clean graphene on suitable substrates. The substrates need to be insulating, ideally have a high dielectric constant, $\kappa$, to prevent gate current leakage and hence allow greater miniaturization. In graphene, the carrier concentration and polarity can be modulated by an electric field. However, the fabrication of graphene devices with an atomically uniform gate dielectric, so as to provide a uniform electric field over the active graphene area without damaging the graphene surface remains a challenge \cite{lee08}. For this reason and the ease with which graphene can be observed, most reported graphene devices are fabricated on Si/SiO$_2$ which forms a global back gate. Attempts to deposit dielectrics on the surface of graphene to form top gates have been made, however the deposition processes employed tend to damage the graphene, see \textit{e.\,g.}\:Ref.\:\onlinecite{lemme07}. The situation for graphene based devices is further complicated due to restrictions imposed by most of the available synthesis routes. Aside from mechanical cleaving, graphene synthesis generally involves the use of metal catalysts \cite{sutter08,reina09,li09} which then necessitates post-synthesis transfer of the graphene or in the case of epitaxially grown graphene in SiC high temperatures are required \cite{emtsev09}. Other synthesis routes like chemical exfoliation or the necessary transfer steps from the aforementioned routes to dielectric substrates introduce high levels of contamination and defects into the graphene making such synthesis routes incompatible with current planar fabrication technology where reproducibility is paramount. Here we report graphene formation over a previously introduced class of catalytically active substrates, metal oxides \cite{rummeli10a,rummeli10b}. We show through transmission electron microscopy (TEM) that graphene nano-flakes can form on MgO and multilayer graphene nano-flakes on both MgO and ZrO$_2$. As high-$\kappa$ dielectrics, these oxides are an attractive natural candidate for graphene based electronics. Furthermore the MgO system can catalyze graphene growth \textit{via} thermal chemical vapor deposition (CVD) at the low temperature of 325\,\textcelsius \cite{rummeli10b}. Growth on ZrO$_2$ can occur at least as low as 480\,\textcelsius.

All reactions were conducted in a CVD oven comprising a quartz tube and a sliding heating element. The samples were placed in an alumina boat. In the case where cyclohexane served as the feedstock, the oven loaded with MgO nano-powder was evacuated to $< 1$\:mbar and heated to a synthesis temperature of 775\,\textcelsius. A pressure of 100\:mbar was used. The reactions were terminated after 17\:s to 300\:s by flushing with argon and sliding the heating element off the reaction zone. Further experiments were performed on MgO and ZrO$_2$ using acetylene as the feedstock. After loading the samples the reactor was evacuated and then exposed to a gentle argon flow (1015\:mbar), and heated to temperatures ranging between 325\,\textcelsius\ and 650\,\textcelsius. The argon flow was held for 10\:min and then switched to a mixture of acetylene and argon. Finally the oven was cooled in an argon atmosphere. In some cases we removed the oxide catalyst by ultrasonication in HCl. The TEM studies were performed on an \mbox{FEI Titan$^3$ 80--300} with an 80\:kV acceleration voltage. The \textit{ab initio} total energy calculations were carried out using density-functional theory (DFT). They were performed within the PBE generalized gradient approximation for the exchange-correlation functional \cite{perdew96} and the projector augmented-wave method \cite{blochl94}, using the Vienna \textit{ab initio} simulation package (\textsc{VASP}) \cite{kresse96}. The activation energies were calculated using the nudged elastic band model as developed by the Henkelman group \cite{henkelman06}. The simulation cell included a slab of four substrate layers with a lateral size of a $4 \times 4$ surface unit cell. This corresponds to 16 atoms on the surface layer. The step surface is prepared by removing eight atoms from the surface layer. In the case of an existing graphene on step edge surface, the lateral size of simulation cell is duplicated. In all simulations the vacuum gap between the slabs was larger than 15\:\AA. The wave functions have been expanded into plane waves up to a kinetic energy cut-off of 400\:eV. The integration in the first Brillouin zone was performed using Monkhorst-Pack grids \cite{monkhorst76} including 25 $k$-points in the irreducible wedge. In all calculations, the two topmost layers of MgO have been optimized and from third layer up fixed in the bulk position until all force components were less than 0.01\:eV/\AA. The convergence of energy differences with respect to the used cut-off energies and $k$-point grids has been tested in all cases within a tolerance of 10\:meV/atom. 

\begin{figure}
  \includegraphics[width=.9\columnwidth]{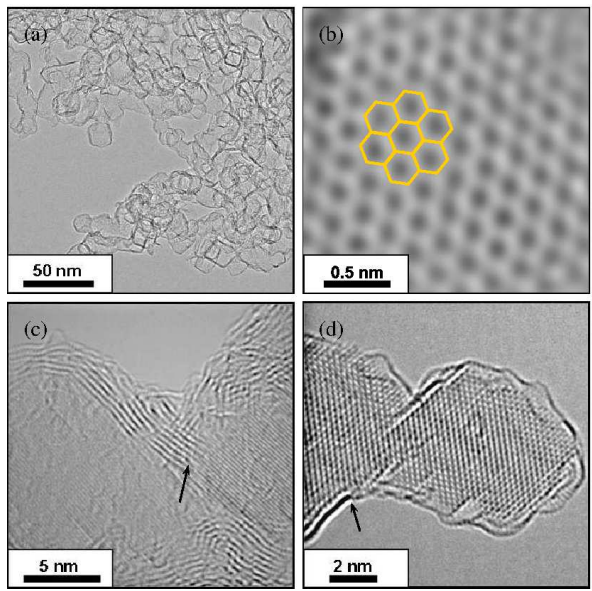}
  \caption{\label{fig_1} (color online) (a)--(c) Multilayer graphene synthesized over MgO from cyclohexane at 775\,\textcelsius\ and 100\:mbar. (a) Few layer graphene shells after removal of MgO core \textit{via} acid treatment. (b) Graphene lattice from single layer shell, (c) graphene layers anchored at step-edges on the (100) surface. (d) Graphene flakes synthesized over MgO from acetylene at 325\,\textcelsius\ and 1015\:mbar. The attachment of graphene layers to the (100) oxide step-edges can be observed. Arrows indicate graphene sheets anchored at step sites.}
\end{figure}
In figure \ref{fig_1}, a variety of micrographs are presented which show we are able to grow graphene flakes on the surface of magnesium oxide and tune the growth from single to multi layers. The samples were grown from cyclohexane at 775\,\textcelsius\ and 100\:mbar. With these conditions, the reaction stops when the catalyst is fully encapsulated by graphitic layers. We found that the number of layers can increase up to nine and can be tuned \textit{via} the reaction time. In figure \ref{fig_1} (c) we highlight the closely spaced parallel lattice fringes of the crystals are those of (100) planes on top of which there is graphitic carbon. Furthermore, as indicated by the arrow, the multi-layers are ubiquitously anchored into step edges on the (100) surfaces. We propose that these step sites initiate the growth of graphitic layers. We argue that these step sites are not only nucleation sites, but also growth sites because growth appears to stop once the particle is fully encapsulated. In this regard, this work corroborates previous studies suggesting the cooperative role of oxide supports in the growth of multi walled carbon nanotubes \cite{rummeli07b}.
\begin{figure}
  \includegraphics[width=.9\columnwidth]{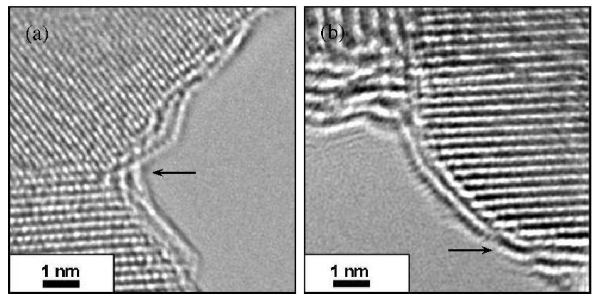}
  \caption{\label{fig_2} Graphitic carbon grown over monoclinic ZrO$_2$ from acetylene at 480\,\textcelsius\ and 1015\:mbar. One clearly sees the anchoring of the graphitic layers onto a step of the (020) crystal face. Arrows indicate graphene sheets anchored at step sites.}
\end{figure}
We also investigated zirconium oxide to further elucidate the atomic surface structure required for graphitic synthesis on oxides. This catalyst has recently been extensively explored by Steiner III \textit{et al.} \cite{steiner09} for the synthesis of carbon nanotubes. The investigators used \textit{in situ} XPS to confirm that zirconium oxide in a monoclinic baddeleyite or slightly oxygen deficient form could form $sp^2$ carbon. In our case, the synthesis of graphitic layers is performed on nano-crystallites which following the reaction are in the baddeleyite form evidenced by the lattice fringes in TEM micrographs of figure \ref{fig_2}. In both cases arrows indicate graphitic shells anchoring on step sites. In Fig.\:\ref{fig_2}(b) it is possible to determine that the graphitic layers are anchored on a face with (020) as its surface normal. 
\begin{figure}
  \includegraphics[width=.9\columnwidth]{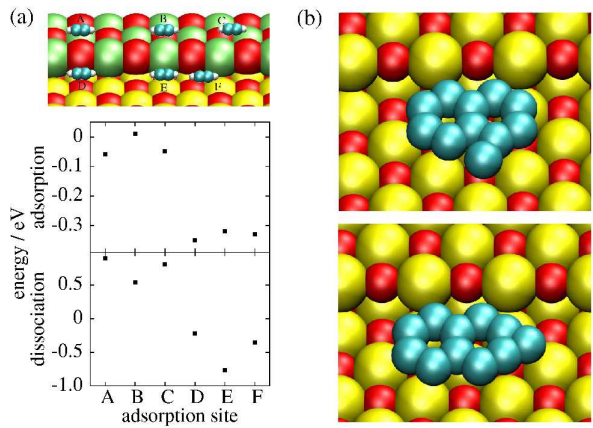}
  \caption{\label{fig_3} (color online) (a) The initial adsorption sites of C$_2$H$_2$ on (100) and step-edge of MgO surface and corresponding calculated adsorption energies for the different sites and dissociative energies of C$_2$H$_2$ to C$_2$H and H. (b) Different configurations of carbon adsorption on a MgO surface with an existing graphene flake. Red\,---\,O, yellow\,---\,Mg, blue\,---\,C, green\,---\,Mg in step plane, and white\,---\,H.}
\end{figure}
The low synthesis temperatures employed suggest the oxides are catalytically active for the formation of $sp^2$ carbon. To support this concept, theoretical studies were conducted on MgO step sites. Based on experimental evidence, we chose a step site on the (100) surface of MgO as a model system for oxide based graphene catalysis. The exact atomic structure of graphene on MgO is undetermined. The lattice constant of MgO is 2.10\:\AA\ and the spatial frequency in the reciprocal lattice of graphene is 2.13\:\AA$^{-1}$. The adsorption and dissociation behavior of C$_2$H$_2$ on (100) and step-edge of MgO surface was studied. Several adsorption sites on the (100) surface and step-edge of MgO are considered as initial adsorption states which are shown in Fig.\:\ref{fig_3}(a). The adsorption energy is defined as the difference between the total energy of the system with the adsorbate and without the adsorbate referred to as the free energy of C$_2$H$_2$. The adsorption energies are shown in Fig.\:\ref{fig_3}(a). In all adsorption sites, the molecule binds very weakly to the (100) surface. The distance from the surface and the center of mass of C$_2$H$_2$ lies between 2.5\:\AA\ and 3\:\AA. These distances are in the physisorption adsorption regime. The dissociative adsorption energy of C$_2$H$_2$ to C$_2$H and H is defined by
\begin{equation*}
  E_{\text{diss.}} = E_{\text{C}_2\text{H}+\text{H}^s} - E_s - E_{\text{C}_2\text{H}_2},
\end{equation*}
where $E_{\text{C}_2\text{H}+\text{H}^s}$ is the total energy of the C$_2$H and the H on the substrate, $E_s$ the total energy of the clean substrate, and $E_{\text{C}_2\text{H}_2}$ the free energy of acetylene in the gas phase. The data shows that the dissociative adsorption of C$_2$H$_2$ on MgO (100) is an endothermic reaction (see Fig.\:\ref{fig_3}). In contrast to the (100) surface, stronger binding is found between C$_2$H$_2$ and the step-edge of MgO. The dissociative adsorption energies indicate exothermic reactions on the step-edge surface. Thus, according to our calculations, the presence of the step-edge on the MgO surface is essential for the decomposition of C$_2$H$_2$. In addition, we have calculated the adsorption of single carbon atoms on several surface positions. For each adsorption position, the lateral position of carbon is fixed to avoid the diffusion of the atom to favorable adsorption sites. The most favorable adsorption site of carbon on an MgO (100) surface is the step-edge adsorption position. In addition, the diffusion barrier of a carbon atom on MgO (100) has been calculated. The total energy of the initial and final configurations have been obtained using geometry optimization. The diffusion barrier is about 0.38\:eV. The diffusion barrier of a hydrogen atom on the surface from on-top of one oxygen atom to the next is 0.2\:eV. Thus, the hydrogen atoms simply diffuse away while the carbon remains on the surface. 

The importance of step sites to the growth of carbon nanotubes is already well established for some metal catalysts. It has been argued elsewhere \cite{helveg04,hofmann05} that the crystalline state of nickel catalyst particles evidences a surface growth mechanism which can divided into four distinct sub-processes\,---\,the adsorbtion of carbon feedstock molecules on the catalyst, dissociation of hydrogen from the precursor, surface diffusion and addition of carbon atoms to the network. Our theoretical investigations, while still preliminary, clearly show each of these subprocesses on MgO (100) step sites (\textit{cf.}\:Fig.\:\ref{fig_3}). The experimental data show graphene nano-flakes anchored at step sites. Questions remain though, for example, what are the effects of the edges of the nano-flakes, their possible interaction with the substrate and the effect of defects in the dielectric. None-the-less, the demonstrated catalytic potential of high-$\kappa$ dielectrics for the catalytic formation of graphene \textit{via} thermal CVD at low temperatures is an advance. 

\medskip
F.B.\:acknowledges the DFG \mbox{(RU 1540/8-1)}, A.B. the A.-v.-Humboldt Stiftung and the BMBF, S.Z.\:the \mbox{IMPRS} ``Dynamical Processes in Atoms, Molecules and Solids'', G.C.\:the South Korean Ministry of Education, Science, and Technology Program, Project WCU ITCE No. R31-2008-000-10100-0., and M.H.R.\:the EU (ECEMP) and the Freistaat Sachsen.

\end{document}